\documentclass[11pt,a4paper]{article}
\usepackage[hyperref]{acl2018}
\usepackage{latexsym}
\usepackage{times}
\usepackage{amsmath}
\usepackage{amssymb}
\usepackage{caption}
\usepackage{graphicx}
\usepackage{url}
\usepackage{multirow}
\usepackage{color}
\usepackage{subfigure}

\aclfinalcopy 

\title{Entity-Duet Neural Ranking: Understanding the Role of Knowledge Graph Semantics in Neural Information Retrieval
}

\author{Zhenghao Liu$^1$ \qquad Chenyan Xiong$^2$ \qquad Maosong Sun$^1$ \thanks{ \ \ Corresponding author: M. Sun (sms@tsinghua.edu.cn)} \qquad Zhiyuan Liu$^1$\\ 
$^1$State Key Laboratory of Intelligent Technology and Systems\\
Beijing National Research Center for Information Science and Technology\\
Department of Computer Science and Technology, Tsinghua University,
Beijing, China \\
$^2$Language Technologies Institute, Carnegie Mellon University\\}

\date{}

\begin{document}
\maketitle


\begin{abstract}
This paper presents the Entity-Duet Neural Ranking Model (\texttt{EDRM}), which introduces knowledge graphs to neural search systems.
\texttt{EDRM} represents queries and documents by their words and entity annotations.
The semantics from knowledge graphs are integrated in the distributed representations of their entities, while the ranking is conducted by interaction-based neural ranking networks. The two components are learned end-to-end, making \texttt{EDRM} a natural combination of entity-oriented search and neural information retrieval.
Our experiments on a commercial search log demonstrate the effectiveness of \texttt{EDRM}.
Our analyses reveal that knowledge graph semantics significantly improve the generalization ability of neural ranking models.
\end{abstract}

\section{Introduction}

The emergence of large scale knowledge graphs has motivated the development of \emph{entity-oriented search}, which utilizes knowledge graphs to improve search engines.
The recent progresses in entity-oriented search include better text representations with entity annotations~\cite{Xiong2016BOE,raviv2016document}, richer ranking features~\cite{daltonentity}, entity-based connections between query and documents~\cite{liu2015latent,EsdRank}, and soft-match query and documents through knowledge graph relations or embeddings~\cite{ESR, SELM}. 
These approaches bring in entities and semantics from knowledge graphs  and have greatly improved the effectiveness of feature-based search systems.

Another frontier of information retrieval is the development of neural ranking models (\emph{neural-IR}).
Deep learning techniques have been used to learn distributed representations of queries and documents that capture their relevance relations (\emph{representation-based})~\cite{cdssm}, or to model the query-document relevancy directly from their word-level interactions (\emph{interaction-based})~\cite{guo2016matching,xiong2017knrm,convknrm}.
Neural-IR approaches, especially the \emph{interaction-based} ones, have greatly improved the ranking accuracy when large scale training data are available~\cite{convknrm}.

Entity-oriented search and neural-IR push the boundary of search engines from two different aspects.
Entity-oriented search incorporates human knowledge from entities and knowledge graph semantics. It has shown promising results on feature-based ranking systems.
On the other hand, neural-IR leverages distributed representations and neural networks to learn more sophisticated ranking models form large-scale training data.
However, it remains unclear how these two approaches interact with each other and 
whether the entity-oriented search has the same advantage in neural-IR methods as in feature-based systems.

This paper explores the role of entities and semantics in neural-IR. We present an Entity-Duet Neural Ranking Model (\texttt{EDRM}) that incorporates entities in interaction-based neural ranking models.
\texttt{EDRM} first learns the distributed representations of entities using their semantics from knowledge graphs: descriptions and types. Then it follows a recent state-of-the-art entity-oriented search framework, the word-entity duet~\cite{xiong2017duet}, and matches documents to queries with both bag-of-words and bag-of-entities.
Instead of manual features, \texttt{EDRM} uses interaction-based neural models~\cite{convknrm} to match query and documents with word-entity duet representations. As a result, \texttt{EDRM} combines entity-oriented search and the interaction based neural-IR; it brings the knowledge graph semantics to neural-IR and enhances entity-oriented search with neural networks.

One advantage of being neural is that \texttt{EDRM} can be learned end-to-end. Given a large amount of user feedback from a commercial search log, the integration of knowledge graph semantics to neural ranker, is learned jointly with the modeling of query-document relevance in \texttt{EDRM}. It provides a convenient data-driven way to leverage external semantics in neural-IR.

Our experiments on a Sogou query log and CN-DBpedia demonstrate the effectiveness of entities and semantics in neural models.
\texttt{EDRM} significantly outperforms the word-interaction-based neural ranking model, \texttt{K-NRM}~\cite{xiong2017duet}, confirming the advantage of entities in enriching word-based ranking.
The comparison with \texttt{Conv-KNRM}~\cite{convknrm}, the recent state-of-the-art neural ranker that models phrase level interactions, provides a more interesting observation: \texttt{Conv-KNRM} predicts user clicks reasonably well, but integrating knowledge graphs using \texttt{EDRM} significantly improves the neural model's generalization ability on more difficult scenarios.

Our analyses further revealed the source of \texttt{EDRM}'s generalization ability: the knowledge graph semantics.
If only treating entities as ids and ignoring their semantics from the knowledge graph, the entity annotations are only a cleaner version of phrases. In neural-IR systems, the embeddings and convolutional neural networks have already done a decent job in modeling phrase-level matches.
However, the knowledge graph semantics brought by \texttt{EDRM} can not yet be captured solely by neural networks; incorporating those human knowledge greatly improves the generalization ability of neural ranking systems.  

\section{Related Work}
Current neural ranking models can be categorized into two groups: representation based and interaction based~\cite{jiafeng2016deep}. The earlier works mainly focus on representation based models. They learn good representations and match them in the learned representation space of query and documents. \texttt{DSSM}~\cite{huang2013learning} and its convolutional version \texttt{CDSSM}~\cite{cdssm} get representations by hashing letter-tri-grams to a low dimension vector. A more recent work uses pseudo-labeling as a weak supervised signal to train the representation based ranking model~\cite{weeksupervise}.

The interaction based models learn word-level interaction patterns from query-document pairs. \texttt{ARC-II}~\cite{arcii} and \texttt{MatchPyramind}~\cite{Pang2016TextMA} utilize Convolutional Neural Network (CNN) to capture complicated patterns from word-level interactions. The Deep Relevance Matching Model (\texttt{DRMM})~\cite{jiafeng2016deep} uses pyramid pooling (histogram) to summarize the word-level similarities into ranking models. \texttt{K-NRM} and \texttt{Conv-KNRM} use kernels to summarize word-level interactions with word embeddings and provide soft match signals for learning to rank. There are also some works establishing position-dependent interactions for ranking models~\cite{pang2017deeprank,hui2017pacrr}. Interaction based models and representation based models can also be combined for further improvements~\cite{www2017duet}.

Recently, large scale knowledge graphs such as DBpedia~\cite{auer2007dbpedia}, Yago~\cite{suchanek2007yago} and Freebase~\cite{bollacker2008freebase} have emerged. Knowledge graphs contain human knowledge about real-word entities and become an opportunity for search system to better understand queries and documents. There are many works focusing on exploring their potential for ad-hoc retrieval. They utilize knowledge as a kind of pseudo relevance feedback corpus~\cite{cao2008selecting} or weight words to better represent query according to well-formed entity descriptions. Entity query feature expansion~\cite{daltontrec} uses related entity attributes as ranking features. 

Another way to utilize knowledge graphs in information retrieval is to build the additional connections from query to documents through related entities. Latent Entity Space (\texttt{LES}) builds an unsupervised model using latent entities' descriptions~\cite{liu2015latent}. \texttt{EsdRank} uses related entities as a latent space, and performs learning to rank with various information retrieval features~\cite{EsdRank}. \texttt{AttR-Duet} develops a four-way interaction to involve cross matches between entity and word representations to catch more semantic relevance patterns~\cite{xiong2017duet}. 

There are many other attempts to integrate knowledge graphs in neural models in related tasks~\cite{Miller2016KeyValueMN,gupta2017entity,kgqa}. 
Our work shares a similar spirit and focuses on exploring the effectiveness of knowledge graph semantics in neural-IR.

\section{Entity-Duet Neural Ranking Model}
\begin{figure*}[tbp]
	\centering
	\includegraphics[width=1.0\linewidth]{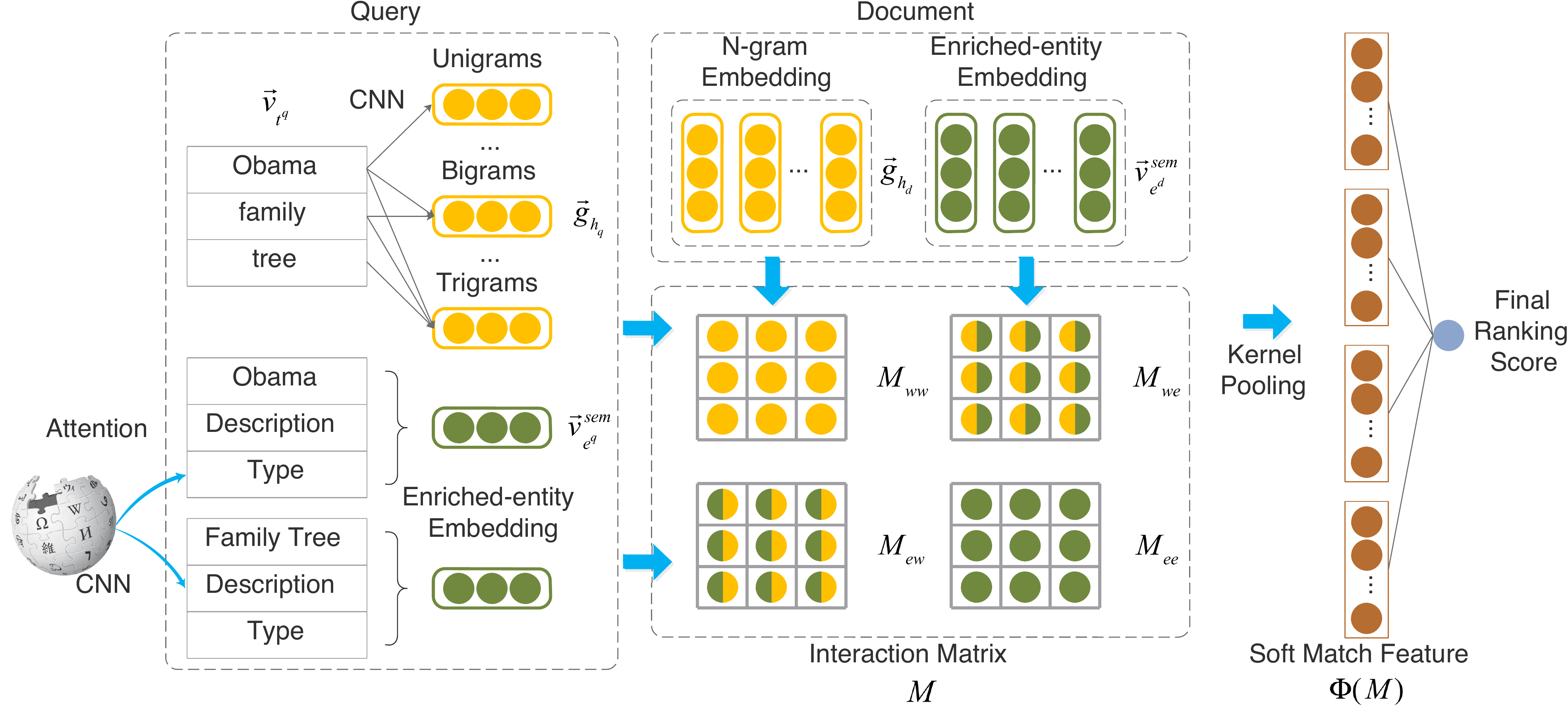}
	\caption{The architecture of EDRM.\label{fig:model}}
\end{figure*}

This section first describes the standard architecture in current interaction based neural ranking models. Then it presents our Entity-Duet Neural Ranking Model, including the semantic entity representation which integrates the knowledge graph semantics, and then the entity-duet ranking framework.
The overall architecture of \texttt{EDRM} is shown in Figure \ref{fig:model}.

\subsection{Interaction based Ranking Models}\label{model.interaction}

Given a query $q$ and a document $d$, interaction based models first build the word-level translation matrix between $q$ and $d$~\cite{berger1999Information}. The translation matrix describes word pairs similarities using word correlations, which are captured by word embedding similarities in interaction based models. 

Typically, interaction based ranking models first map each word $t$ in $q$ and $d$ to an $L$-dimensional embedding $\vec{v}_{t}$ with an embedding layer $\text{Emb}_w$:
\begin{equation}
\small
\vec{v}_{t} = \text{Emb}_w(t).
\end{equation}

It then constructs the interaction matrix $M$ based on query and document embeddings. Each element $M^{ij}$ in the matrix, compares the $i$th word in $q$ and the $j$th word in $d$, e.g. using the cosine similarity of word embeddings:
\begin{equation}
\small
M^{ij} = \cos (\vec{v}_{t_i^q}, \vec{v}_{t_j^d}).
\end{equation}

With the translation matrix describing the term level matches between query and documents, the next step is to calculate the final ranking score from the matrix.
Many approaches have been developed in interaction base neural ranking models, but in general, that would include a feature extractor $\phi()$ on $M$ and then one or several ranking layers to combine the features to the ranking score.

\subsection{Semantic Entity Representation}
\texttt{EDRM} incorporates the semantic information about an entity from the knowledge graphs into its representation. The representation includes three embeddings: entity embedding, description embedding, and type embedding, all in $L$ dimension and are combined to generate the semantic representation of the entity.

\textbf{Entity Embedding} uses an $L$-dimensional embedding layer $\text{Emb}_e$ to get an entity embedding $\vec{v}_{e}^{\text{emb}}$ for $e$:
\begin{equation}
\small
\begin{aligned}
\vec{v}_{e}^{\text{emb}} = \text{Emb}_e(e).
\end{aligned}
\end{equation}

\textbf{Description Embedding} encodes an entity description which contains $m$ words and explains the entity. \texttt{EDRM} first employs the word embedding layer $\text{Emb}_w$ to embed the description word $w$ to $\vec{v}_{w}$. Then it combines all embeddings in text to an embedding matrix $\vec{V}_{w}$. Next, it leverages convolutional filters to slide over the text and compose the $h$ length n-gram as $\vec{g}_{e}^{j}$:
\begin{equation}
\small
\vec{g}_{e}^{j} =  \text{ReLu} (W_{\text{CNN}} \cdot \vec{V}_{w}^{j:j+h} + \vec{b}_{\text{CNN}}),
\end{equation}
where $W_{\text{CNN}}$ and $\vec{b}_{\text{CNN}}$ are two parameters of the covolutional filter. 

Then we use max pooling after the convolution layer to generate the description embedding  ${\vec{v}^{\text{des}}_{e}}$:
\begin{equation}
\small
\vec{v}^{\text{des}}_{e} = \max (\vec{g}_{e}^1 ,..., \vec{g}_{e}^j ,..., \vec{g}_{e}^m).
\end{equation}

\textbf{Type Embedding} encodes the categories of entities. Each entity $e$ has $n$ kinds of types $F_{e} = \{ f_1, ..., f_j, ..., f_n\}$. \texttt{EDRM} first gets the $f_j$ embedding $\vec{v}_{f_j}$ through the type embedding layer $\text{Emb}_\text{tp}$: 
\begin{equation}
\small
\begin{aligned}
\vec{v}_{f_j}^{\text{emb}} = \text{Emb}_\text{tp}(e).
\end{aligned}
\end{equation}

Then \texttt{EDRM} utilizes an attention mechanism to combine entity types to the type embedding $\vec{v}_{e}^{\text{type}}$:
\begin{equation}
\small
\vec{v}_{e}^{\text{type}} = \sum_j^n a_j  \vec{v}_{f_j},
\end{equation}
where $a_j$ is the attention score, calculated as:
\begin{equation}
\small
a_j = \frac{\exp (P_j)}{\sum_l^n \exp(P_l )},
\end{equation}
\begin{equation}
\small
P_j = (\sum_i W_{\text{bow}} \vec{v}_{t_i}) \cdot \vec{v}_{f_j}.
\end{equation}
$P_j$ is the dot product of the query or document representation and type embedding $f_j$. We leverage bag-of-words for query or document encoding. $W_{\text{bow}}$ is a parameter matrix. 

\textbf{Combination.} 
The three embeddings are combined by an linear layer to generate the semantic representation of the entity:
\begin{equation}
\small
\vec{v}_{e}^{\text{sem}} = \vec{v}_{e}^{\text{emb}} +  W_{e} (\vec{v}_{e}^{\text{des}} \oplus \vec{v}_{e}^{\text{type}})^T + \vec{b}_{e}.
\end{equation}
$W_{e}$ is an $L \times 2L$ matrix and $\vec{b}_{e}$ is an $L$-dimensional vector.

\subsection{Neural Entity-Duet Framework}

Word-entity duet~\cite{xiong2017duet} is a recently developed framework in entity-oriented search. It utilizes the duet representation of bag-of-words and bag-of-entities to match $q$-$d$ with hand crafted features. This work introduces it to neural-IR.

We first construct bag-of-entities $q^e$ and $d^e$ with entity annotation as well as bag-of-words $q^w$ and $d^w$ for $q$ and $d$. The duet utilizes a four-way interaction: query words to document words ($q^w$-$d^w$), query words to documents entities ($q^w$-$d^e$), query entities to document words ($q^e$-$d^w$) and query entities to document entities ($q^e$-$d^e$). 

Instead of features, \texttt{EDRM} uses a translation layer that calculates similarity between a pair of query-document terms:
($\vec{v}_{w^q}^{i}$ or $\vec{v}_{e^q}^{i}$) and ($\vec{v}_{w^d}^{j}$ or $\vec{v}_{e^d}^{j}$). It constructs the interaction matrix $M = \{M_{ww}, M_{we}, M_{ew}, M_{ee}\}$. And $M_{ww}, M_{we}, M_{ew}, M_{ee}$ denote interactions of $q^w$-$d^w$, $q^w$-$d^e$, $q^e$-$d^w$, $q^e$-$d^e$ respectively. And elements in them are the cosine similarities of corresponding terms:
\begin{equation}
\small
\begin{aligned}
M_{ww}^{ij} = \cos (\vec{v}_{w^q}^{i}, \vec{v}_{w^d}^{j})&;  M_{ee}^{ij} = \cos (\vec{v}_{e^q}^{i}, \vec{v}_{e^d}^{j})\\
M_{ew}^{ij} = \cos (\vec{v}_{e^q}^{i}, \vec{v}_{w^d}^{j})&;  M_{we}^{ij} = \cos (\vec{v}_{w^q}^{i}, \vec{v}_{e^d}^{j}).
\end{aligned}
\end{equation}

The final ranking feature $\Phi(\mathcal{M})$ is a concatenation ($\oplus$) of four cross matches ($\phi(M)$):

\begin{equation}
\small
\Phi(\mathcal{M}) = \phi(M_{ww}) \oplus  \phi(M_{we}) \oplus  \phi(M_{ew}) \oplus  \phi(M_{ee}),
\end{equation}
where the $\phi$ can be any function used in interaction based neural ranking models.

The entity-duet presents an effective way to cross match query and document in entity and word spaces. In \texttt{EDRM}, it introduces the knowledge graph semantics representations into neural-IR models.

\section{Integration with Kernel based Neural Ranking Models}
The duet translation matrices provided by \texttt{EDRM} can be plugged into any standard interaction based neural ranking models.
This section expounds special cases where it is integrated with \texttt{K-NRM}~\cite{xiong2017knrm} and  \texttt{Conv-KNRM}~\cite{convknrm}, two recent state-of-the-arts. 

\texttt{K-NRM} uses $K$ Gaussian kernels to extract the matching feature $\phi(M)$ from the translation matrix $M$. Each kernel $K_k$ summarizes the translation scores as soft-TF counts, generating a $K$-dimensional feature vector $\phi(M) = \{ K_1(M), ... ,K_K(M) \}$:
\begin{equation}
\small
K_k(M) = \sum_j \exp (- \frac{M^{ij}-\mu_k}{2 \delta_k^2}).
\end{equation}
$\mu_k$ and $\delta_k$ are the mean and width for the $k$th kernel. \texttt{Conv-KNRM} extend \texttt{K-NRM} incorporating $h$-gram compositions $\vec{g}_h^i$ from text embedding $\vec{V}_{T}$ using CNN:
\begin{equation}
\small
\vec{g}_h^i = \text{relu} (W_h \cdot \vec{V}_{T}^{i:i+h} + \vec{v}_h).
\end{equation}

Then a translation matrix $M_{h_q,h_d}$ is constructed. Its elements are the similarity scores of $h$-gram pairs between query and document:
\begin{equation}
\small
M_{h_q,h_d} = \cos (\vec{g}_{h_q}^i, \vec{g}_{h_d}^j).
\end{equation}

We also extend word n-gram cross matches to word entity duet matches:
\begin{equation}
\small
\Phi(\mathcal{M}) = \phi(M_{1,1}) \oplus ... \oplus \phi(M_{h_q,h_d}) \oplus ... \oplus \phi (M_{ee}).
\end{equation}

Each ranking feature $\phi(M_{h_q,h_d})$ contains three parts: query $h_q$-gram and document $h_d$-gram match feature ($\phi(M_{{ww}^{h_q,h_d}})$), query entity and document $h_d$-gram match feature ($\phi(M_{{ew}^{1,h_d}})$), and query $h_q$-gram and document entity match feature ($\phi(M_{{ww}^{h_q,1}})$):
\begin{equation}
\small
\phi(M_{h_q,h_d}) = \phi(M_{{ww}^{h_q,h_d}}) \oplus \phi(M_{{ew}^{1,h_d}})  \oplus \phi(M_{{we}^{h_q,1}}).
\end{equation}

We then use learning to rank to combine ranking feature $\Phi(\mathcal{M})$ to produce the final ranking score:
\begin{equation}
\small
f(q,d) = \text{tanh} (\omega_r^T \Phi(\mathcal{M}) + b_r).
\end{equation}
$\omega_r$ and $b_r$ are the ranking parameters. $\text{tanh}$ is the activation function. 

We use standard pairwise loss to train the model:
\begin{equation}
\small
l = \sum_q \sum_{d^+,d^- \in D_q^{+,-}} \max (0, 1 - f(q,d^+) + f(q,d^-)),
\end{equation}
where the $d^+$ is a document ranks higher than $d^-$. 

With sufficient training data, the whole model is optimized end-to-end with back-propagation. During the process, the integration of the knowledge graph semantics, entity embedding, description embeddings, type embeddings, and matching with entities-are learned jointly with the ranking neural network.

\section{Experimental Methodology}\label{section.experiment}

This section describes the dataset, evaluation metrics, knowledge graph,  baselines, and implementation details of our experiments.

\textbf{Dataset.} Our experiments use a query log from Sogou.com, a major Chinese searching engine~\cite{sogou}. The exact same dataset and training-testing splits in the previous research \cite{xiong2017knrm,convknrm} are used. 
They defined the ad-hoc ranking task in this dataset as re-ranking the candidate documents provided by the search engine. All Chinese texts are segmented by ICTCLAS~\cite{ICTCLAS}, after that they are treated the same as English.

Prior research leverages clicks to model user behaviors and infer reliable relevance signals using click models \cite{chuklin2015click}. DCTR and TACM are two click models: DCTR calculates the relevance scores of a query-document pair based on their click through rates (CTR); TACM~\cite{tacm} is a more sophisticated model that uses both clicks and dwell times. Following previous research~\cite{xiong2017knrm}, both DCTR and TACM are used to infer labels. DCTR inferred relevance labels are used in training. Three testing scenarios are used: Testing-SAME, Testing-DIFF and Testing-RAW.

Testing-SAME uses DCTR labels, the same as in training. Testing-DIFF evaluates models performance based on TACM inferred relevance labels. Testing-RAW evaluates ranking models through user clicks, which tests ranking performance for the most satisfying document. Testing-DIFF and Testing-RAW are harder scenarios that challenge the generalization ability of all models, because their training labels and testing labels are generated differently~\cite{xiong2017knrm}.

\begin{figure}[t]
	\centering
	\subfigure[Statistic of queries]{
		\label{fig:distribution:a}
		\includegraphics[width=0.48\linewidth]{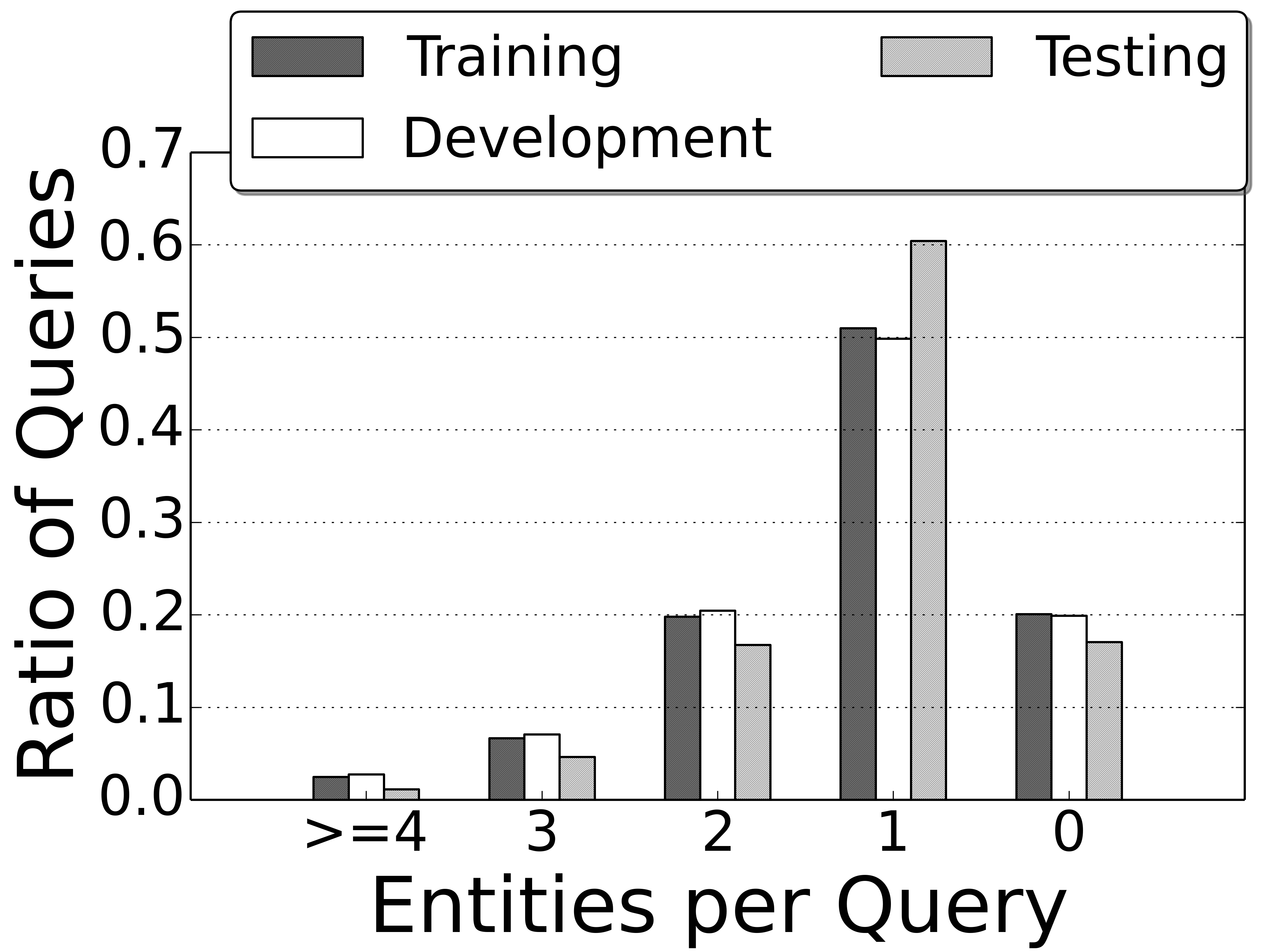}}
	\subfigure[Statistic of documents]{
		\label{fig:distribution:b}
		\includegraphics[width=0.48\linewidth]{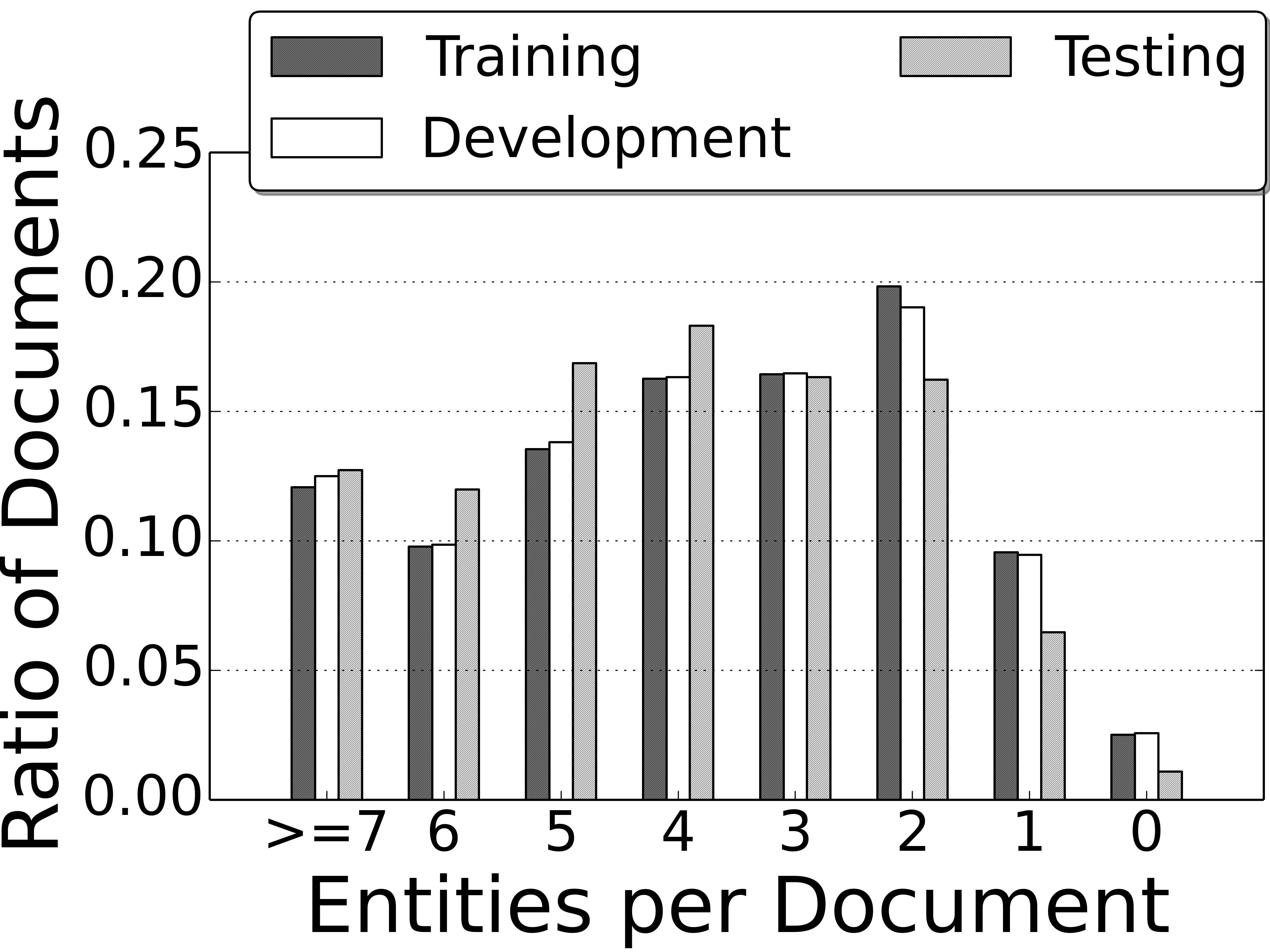}}
	\caption{Query and document distributions. Queries and documents are grouped by the number of entities.}\label{fig:distribution}
\end{figure}
\begin{table*}[h]
	\centering
	\caption{Ranking accuracy of EDRM-KNRM, EDRM-CKNRM and baseline methods. Relative performances compared with K-NRM are in percentages. $\dagger$, $\ddagger$, $\mathsection$, $\mathparagraph$, $*$ indicate statistically significant improvements over DRMM$^{\dagger}$, CDSSM$^{\ddagger}$, MP$^{\mathsection}$, K-NRM$^{\mathparagraph}$ and Conv-KNRM$^{*}$ respectively.}\label{table.overall_acc}
	\resizebox{\textwidth}{23mm}{
		\begin{tabular}{l|lr|lr|lr|lr|lr}
			\hline
			&	\multicolumn{4}{c|}{\textbf{Testing-SAME}}	&	\multicolumn{4}{c|}{\textbf{Testing-DIFF}} & \multicolumn{2}{c}{\textbf{Testing-RAW}} \\ \hline
			\textbf{Method}	&	\multicolumn{2}{c|}{\textbf{NDCG@1}}	&	\multicolumn{2}{c|}{\textbf{NDCG@10}}   &	\multicolumn{2}{c|}{\textbf{NDCG@1}}	&	\multicolumn{2}{c|}{\textbf{NDCG@10}} & \multicolumn{2}{c}{\textbf{MRR}}\\ \hline
			\texttt{BM25}	
			& ${0.1422}$ & $ -46.24\%  $
			& ${0.2868}$ & $ -31.67\%  $
			& ${0.1631}$ & $ -45.63\%  $
			& ${0.3254}$ & $ -23.04\%  $
			& ${0.2280}$ & $ -33.86\%  $  \\
			
			\texttt{RankSVM}	
			& ${0.1457}$ & $ -44.91\%  $
			& ${0.3087}$ & $ -26.45\%  $
			& ${0.1700}$ & $ -43.33\%  $
			& ${0.3519}$ & $ -16.77\%  $
			& ${0.2241}$ & $ -34.99\%  $  \\

			\texttt{Coor-Ascent}	
			& ${0.1594}$ & $ -39.74\%  $
			& ${0.3547}$ & $ -15.49\%  $
			& ${0.2089}$ & $ -30.37\%  $
			& ${0.3775}$ & $ -10.71\%  $
			& ${0.2415}$ & $ -29.94\%  $  \\ \hline

			\texttt{DRMM}
			& ${0.1367}$ & $ -48.34\%  $
			& ${0.3134}$ & $ -25.34\%  $
			& ${0.2126}^{\ddagger }$ & $ -29.14\%  $
			& ${0.3592}^{\mathsection }$ & $ -15.05\%  $
			& ${0.2335}$ & $ -32.26\%  $  \\

			\texttt{CDSSM}	
			& ${0.1441}$ & $ -45.53\%  $
			& ${0.3329}$ & $ -20.69\%  $
			& ${0.1834}$ & $ -38.86\%  $
			& ${0.3534}$ & $ -16.41\%  $
			& ${0.2310}$ & $ -33.00\%  $  \\

			\texttt{MP}	
			& ${0.2184}^{\dagger \ddagger }$ & $ -17.44\%  $
			& ${0.3792}^{\dagger \ddagger }$ & $ -9.67\%  $
			& ${0.1969}$ & $ -34.37\%  $
			& ${0.3450}$ & $ -18.40\%  $
			& ${0.2404}$ & $ -30.27\%  $ \\
			
			\texttt{K-NRM}	
			& $0.2645$ & --  
			& $0.4197$ & --
			& $0.3000$ & --  
			& $0.4228$ & --
			& $0.3447$ & --   \\

			\texttt{Conv-KNRM}	
			& ${0.3357}^{\dagger \ddagger \mathsection \mathparagraph }$ & $ +26.90\%  $
			& ${0.4810}^{\dagger \ddagger \mathsection \mathparagraph }$ & $ +14.59\%  $
			& ${0.3384}^{\dagger \ddagger \mathsection \mathparagraph }$ & $ +12.81\%  $
			& ${0.4318}^{\dagger \ddagger \mathsection }$ & $ +2.14\%  $
			& ${0.3582}^{\dagger \ddagger \mathsection }$ & $ +3.91\%  $\\
			\hline
			
			\texttt{EDRM-KNRM}	
			& ${0.3096}^{\dagger \ddagger \mathsection \mathparagraph }$ & $ +17.04\%  $
			& ${0.4547}^{\dagger \ddagger \mathsection \mathparagraph }$ & $ +8.32\%  $
			& ${0.3327}^{\dagger \ddagger \mathsection \mathparagraph }$ & $ +10.92\%  $
			& ${0.4341}^{\dagger \ddagger \mathsection \mathparagraph }$ & $ +2.68\%  $
			& ${0.3616}^{\dagger \ddagger \mathsection \mathparagraph }$ & $ +4.90\% $  \\

			\texttt{EDRM-CKNRM}	
			& $\textbf{0.3397}^{\dagger \ddagger \mathsection \mathparagraph }$ & $ +28.42\%  $
			& $\textbf{0.4821}^{\dagger \ddagger \mathsection \mathparagraph }$ & $ +14.86\%  $
			& $\textbf{0.3708}^{\dagger \ddagger \mathsection \mathparagraph * }$ & $ +23.60\%  $
			& $\textbf{0.4513}^{\dagger \ddagger \mathsection \mathparagraph * }$ & $ +6.74\%  $
			& $\textbf{0.3892}^{\dagger \ddagger \mathsection \mathparagraph * }$ & $ +12.90\%  $ \\
			
			\hline 
	\end{tabular}}
\end{table*}

\textbf{Evaluation Metrics.} NDCG@1 and NDCG@10 are used in Testing-SAME and Testing-DIFF. MRR is used for Testing-Raw. Statistic significances are tested by permutation test with P$<0.05$. All are the same as in previous research~\cite{xiong2017knrm}. 

\textbf{Knowledge Graph.} 
We use CN-DBpedia~\cite{xu2017cn}, a large scale Chinese knowledge graph based on Baidu Baike, Hudong Baike, and Chinese Wikipedia. CN-DBpedia contains 10,341,196 entities and 88,454,264 relations. 
The query and document entities are annotated by CMNS, the commonness (popularity) based entity linker~\cite{hasibi2017entity}.
CN-DBpedia and CMNS provide good coverage on our queries and documents. As shown in Figure~\ref{fig:distribution}, the majority of queries have at least one entity annotation; the average number of entity annotated per document title is about four.

\textbf{Baselines.}
The baselines include feature-based ranking models and neural ranking models. Most of the baselines are borrowed from previous research~\cite{xiong2017knrm,convknrm}. 

\emph{Feature-based} baselines include two learning to rank systems, \texttt{RankSVM} \cite{ranksvm} and coordinate ascent (\texttt{Coor-Accent}) \cite{coorascent}. The standard word-based unsupervised retrieval model, \texttt{BM25}, is also compared.

\emph{Neural} baselines include \texttt{CDSSM} \cite{cdssm}, MatchPyramid (\texttt{MP}) \cite{Pang2016TextMA}, \texttt{DRMM}  \cite{grauman2005pyramid}, \texttt{K-NRM} \cite{xiong2017knrm} and \texttt{Conv-KNRM} \cite{convknrm}. \texttt{CDSSM} is representation based. It uses CNN to build query and document representations on word letter-tri-grams (or Chinese characters). \texttt{MP} and \texttt{DRMM} are both interaction based models. They use CNNs or histogram pooling to extract features from embedding based translation matrix. 

Our main baselines are \texttt{K-NRM} and \texttt{Conv-KNRM}, the recent state-of-the-art neural models on the Sogou-Log dataset. The goal of our experiments is to explore the effectiveness of knowledge graphs in these state-of-the-art interaction based neural models.

\textbf{Implementation Details.} The dimension of word embedding, entity embedding and type embedding are 300. Vocabulary size of entities and words are 44,930 and 165,877. \texttt{Conv-KNRM} uses one layer CNN with 128 filter size for the n-gram composition. Entity description encoder is a one layer CNN with 128 and 300 filter size for \texttt{Conv-KNRM} and \texttt{K-NRM} respectively.

All models are implemented with PyTorch. Adam is utilized to optimize all parameters with learning rate $=$ 0.001, $\epsilon=1e-5$ and early stopping with the practice of 5 epochs. 

There are two versions of \texttt{EDRM}: \texttt{EDRM-KNRM} and \texttt{EDRM-CKNRM}, integrating with \texttt{K-NRM} and \texttt{Conv-KNRM} respectively. The first one (\texttt{K-NRM}) enriches the word based neural ranking model with entities and knowledge graph semantics; the second one (\texttt{Conv-KNRM}) enriches the n-gram based neural ranking model.

\begin{table*}[t]
	\centering
	\caption{Ranking accuracy of adding diverse semantics based on K-NRM and Conv-KNRM. Relative performances compared are in percentages. $\dagger$, $\ddagger$, $\mathsection$, $\mathparagraph$, $*$, $**$ indicate statistically significant improvements over K-NRM$^{\dagger}$ (or Conv-KNRM$^{\dagger}$), +Embed$^{\ddagger}$, +Type$^{\mathsection}$, +Description$^{\mathparagraph}$, +Embed+Type$^{*}$ and +Embed+Description$^{**}$ respectively.}
	\label{table.component}
	\resizebox{\textwidth}{29mm}{
		\begin{tabular}{l|lr|lr|lr|lr|lr}
			\hline
			&	\multicolumn{4}{c|}{\textbf{Testing-SAME}}	&	\multicolumn{4}{c|}{\textbf{Testing-DIFF}} & \multicolumn{2}{c}{\textbf{Testing-RAW}} \\ \hline
			\textbf{Method}	&	\multicolumn{2}{c|}{\textbf{NDCG@1}}	&	\multicolumn{2}{c|}{\textbf{NDCG@10}}   &	\multicolumn{2}{c|}{\textbf{NDCG@1}}	&	\multicolumn{2}{c|}{\textbf{NDCG@10}} & \multicolumn{2}{c}{\textbf{MRR}}\\ \hline
			\texttt{K-NRM}	
			& $0.2645$ & --  
			& $0.4197$ & --
			& $0.3000$ & --
			& $0.4228$ & --
			& $0.3447$ & --  \\
			
			\hline
			
			\texttt{+Embed}	
			& ${0.2743}$ & $ +3.68\%  $
			& ${0.4296}$ & $ +2.35\%  $
			& ${0.3134}$ & $ +4.48\%  $
			& ${0.4306}$ & $ +1.86\%  $
			& ${0.3641}^{\dagger }$ & $ +5.62\%  $  \\
			
			\texttt{+Type}	
			& ${0.2709}$ & $ +2.41\%  $
			& ${0.4395}^{\dagger }$ & $ +4.71\%  $
			& ${0.3126}$ & $ +4.20\%  $
			& ${0.4373}^{\dagger }$ & $ +3.43\%  $
			& ${0.3531}$ & $ +2.43\%  $  \\
			
			\texttt{+Description}	
			& ${0.2827}$ & $ +6.86\%  $
			& ${0.4364}^{\dagger }$ & $ +3.97\%  $
			& ${0.3181}$ & $ +6.04\%  $
			& ${0.4306}$ & $ +1.86\%  $
			& $\textbf{0.3691}^{\dagger \mathsection * }$ & $ +7.06\%  $  \\
			
			\texttt{+Embed+Type}	
			& ${0.2924}^{\dagger }$ & $ +10.52\%  $
			& ${0.4533}^{\dagger \ddagger \mathsection \mathparagraph }$ & $ +8.00\%  $
			& ${0.3034}$ & $ +1.13\%  $
			& ${0.4297}$ & $ +1.65\%  $
			& ${0.3544}$ & $ +2.79\%  $  \\
			
			\texttt{+Embed+Description}	
			& ${0.2891}$ & $ +9.29\%  $
			& ${0.4443}^{\dagger \ddagger }$ & $ +5.85\%  $
			& ${0.3197}$ & $ +6.57\%  $
			& ${0.4304}$ & $ +1.80\%  $
			& ${0.3564}$ & $ +3.38\%  $  \\
			
			\texttt{Full Model}	
			& $\textbf{0.3096}^{\dagger \ddagger \mathsection }$ & $ +17.04\%  $
			& $\textbf{0.4547}^{\dagger \ddagger \mathsection \mathparagraph }$ & $ +8.32\%  $
			& $\textbf{0.3327}^{\dagger * }$ & $ +10.92\%  $
			& $\textbf{0.4341}^{\dagger }$ & $ +2.68\%  $
			& ${0.3616}^{\dagger }$ & $ +4.90\%  $ \\
			\hline

			\texttt{Conv-KNRM}	
			& ${0.3357}$ & --
			& ${0.4810}$ & --
			& ${0.3384}$ & --
			& ${0.4318}$ & --
			& ${0.3582}$ & --  \\
			
			\hline
			
			\texttt{+Embed}	
			& ${0.3382}$ & $ +0.74\%  $
			& $\textbf{0.4831}$ & $ +0.44\%  $
			& ${0.3450}$ & $ +1.94\%  $
			& ${0.4413}$ & $ +2.20\%  $
			& ${0.3758}^{\dagger }$ & $ +4.91\%  $  \\
			\texttt{+Type}	
			& ${0.3370}$ & $ +0.38\%  $
			& ${0.4762}$ & $ -0.99\%  $
			& ${0.3422}$ & $ +1.12\%  $
			& ${0.4423}^{\dagger }$ & $ +2.42\%  $
			& ${0.3798}^{\dagger }$ & $ +6.02\%  $  \\
			
			\texttt{+Description}	
			& ${0.3396}$ & $ +1.15\%  $
			& ${0.4807}$ & $ -0.05\%  $
			& ${0.3533}$ & $ +4.41\%  $
			& ${0.4468}^{\dagger }$ & $ +3.47\%  $
			& ${0.3819}^{\dagger }$ & $ +6.61\%  $  \\

			\texttt{+Embed+Type}	
			& $\textbf{0.3420}$ & $ +1.88\%  $
			& ${0.4828}$ & $ +0.39\%  $
			& ${0.3546}$ & $ +4.79\%  $
			& ${0.4491}^{\dagger }$ & $ +4.00\%  $
			& ${0.3805}^{\dagger }$ & $ +6.22\%  $  \\
			
			\texttt{+Embed+Description}	
			& ${0.3382}$ & $ +0.73\%  $
			& ${0.4805}$ & $ -0.09\%  $
			& ${0.3608}$ & $ +6.60\%  $
			& ${0.4494}^{\dagger }$ & $ +4.08\%  $
			& ${0.3868}^{\dagger }$ & $ +7.99\%  $  \\
			
			\texttt{Full Model}	
			& ${0.3397}$ & $ +1.19\%  $
			& ${0.4821}$ & $ +0.24\%  $
			& $\textbf{0.3708}^{\dagger \ddagger \mathsection }$ & $ +9.57\%  $
			& $\textbf{0.4513}^{\dagger \ddagger }$ & $ +4.51\%  $
			& $\textbf{0.3892}^{\dagger \ddagger }$ & $ +8.65\%  $ \\
			\hline
			
	\end{tabular}}
\end{table*}

\begin{figure*}[h]
	\centering
	\subfigure[Kernel weight distribution for EDRM-KNRM.\label{fig:overall_contribution:EDKNRM}]{
		\includegraphics[width=0.48\linewidth]{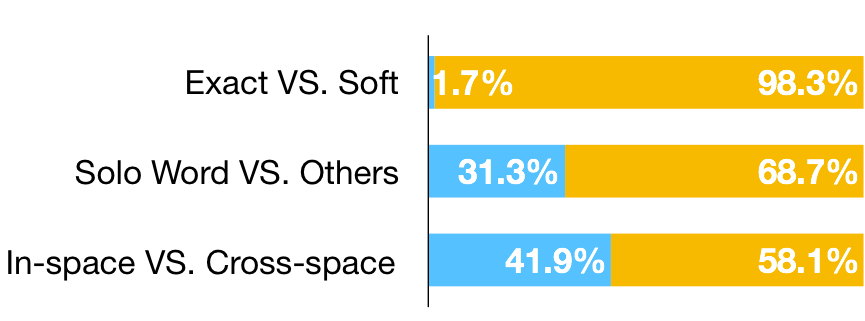}}
	\subfigure[Kernel weight distribution for EDRM-CKNRM.\label{fig:overall_contribution:EDCKNRM}]{
		\includegraphics[width=0.48\linewidth]{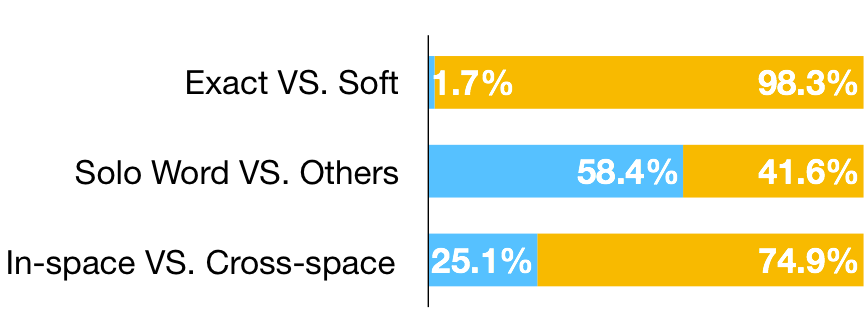}}   
	\caption{Ranking contribution for EDRM. Three scenarios are presented: Exact VS. Soft compares the weights of exact match kernel and others; Solo Word VS. Others shows the proportion of only text based matches; In-space VS. Cross-space compares in-space matches and cross-space matches.}\label{fig:overall_contribution}
\end{figure*}

\section{Evaluation Results}\label{section.evaluation}
Four experiments are conducted to study the effectiveness of \texttt{EDRM}: the overall performance, the contributions of matching kernels, the ablation study, and the influence of entities in different scenarios. We also do case studies to show effect of \texttt{EDRM} on document ranking. 

\subsection{Ranking Accuracy}\label{evaluation.overall}
The ranking accuracies of the ranking methods are shown in Table \ref{table.overall_acc}. \texttt{K-NRM} and \texttt{Conv-KNRM} outperform other baselines in all testing scenarios by large margins as shown in previous research.

\texttt{EDRM-KNRM} out performs \texttt{K-NRM} by over 10\% improvement in Testing-SAME and Testing-DIFF. \texttt{EDRM-CKNRM} has almost same performance on Testing-SAME with \texttt{Conv-KNRM}. A possible reason is that, entity annotations provide effective phrase matches, but \texttt{Conv-KNRM} is also able to learn phrases matches automatically from data. However, \texttt{EDRM-CKNRM} has significant improvement on Testing-DIFF and Testing-RAW. Those demonstrate that \texttt{EDRM} has strong ability to overcome domain differences from different labels.

These results show the effectiveness and the generalization ability of \texttt{EDRM}. In the following experiments, we study the source of this generalization ability.

\subsection{Contributions of Matching Kernels}
This experiment studies the contribution of knowledge graph semantics by investigating the weights learned on the different types of matching kernels.

As shown in Figure \ref{fig:overall_contribution:EDKNRM}, most of the weight in \texttt{EDRM-KNRM} goes to soft match (Exact VS. Soft); entity related matches play an as important role as word based matches (Solo Word VS. Others); cross-space matches are more important than in-space matches (In-space VS. Cross-space). As shown in Figure \ref{fig:overall_contribution:EDCKNRM}, the percentages of word based matches and cross-space matches are more important in \texttt{EDRM-CKNRM} compared to in \texttt{EDRM-KNRM}. 

\begin{figure}[t]
	\includegraphics[width=\linewidth]{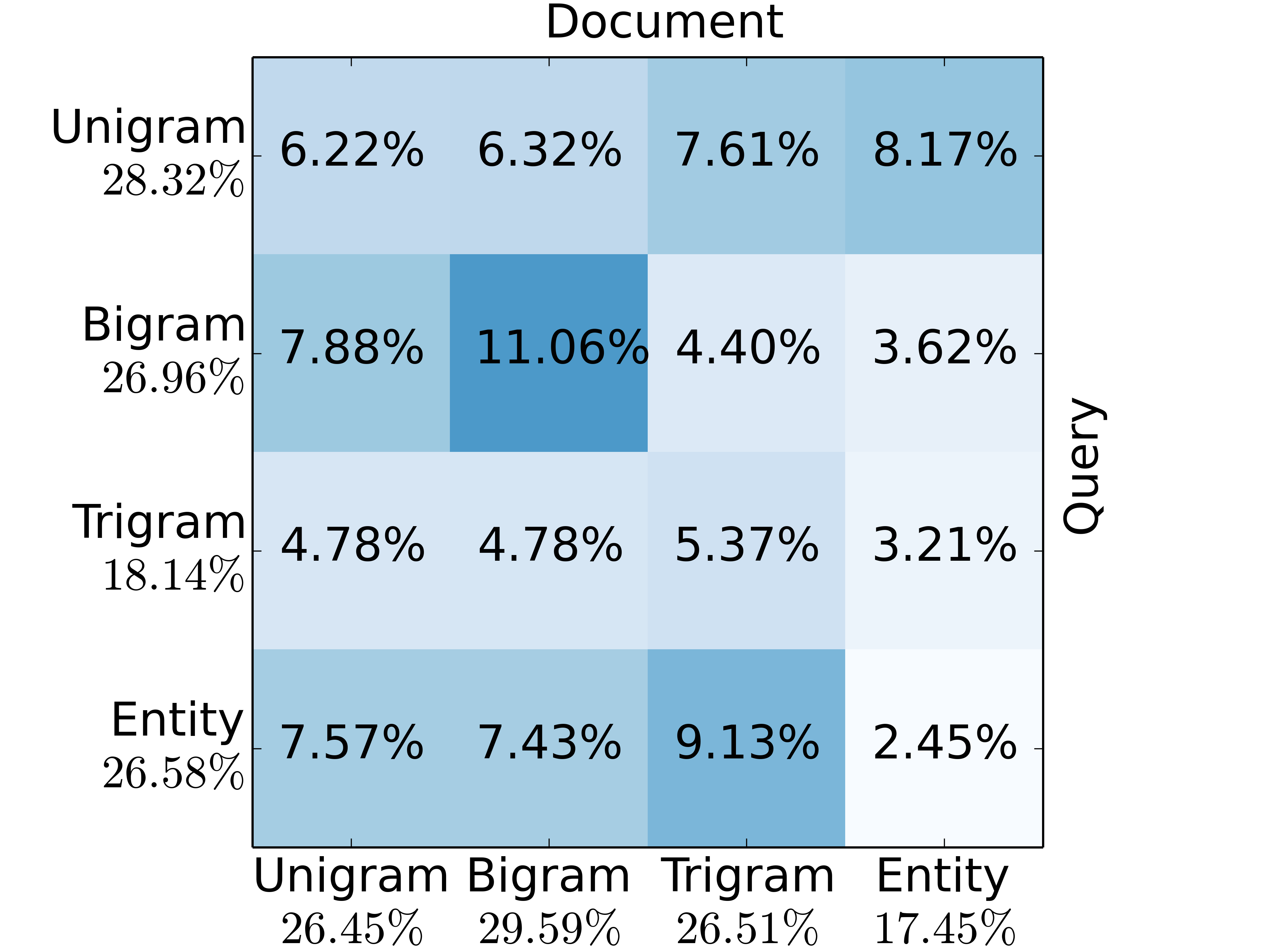}  
	\caption{Individual kernel weight for EDRM-CKNRM. X-axis and y-axis denote document and query respectively.}\label{fig:individual_contribution}
\end{figure}

The contribution of each individual match type in \texttt{EDRM-CKNRM} is shown in Figure \ref{fig:individual_contribution}. The weight of unigram, bigram, trigram, and entity is almost uniformly distributed, indicating the effectiveness of entities and all components are important in \texttt{EDRM-CKNRM}.

\subsection{Ablation Study}
This experiment studies which part of the knowledge graph semantics leads to the effectiveness and generalization ability of \texttt{EDRM}. 

There are three types of embeddings incorporating different aspects of knowledge graph information:
entity embedding (Embed), description embedding (Description) and type embedding (Type). This experiment starts with the word-only \texttt{K-NRM} and \texttt{Conv-KNRM}, and adds these three types of embedding individually or two-by-two (Embed+Type and Embed+Description).

The performances of \texttt{EDRM} with different groups of embeddings are shown in Table~\ref{table.component}. The description embeddings show the greatest improvement among the three embeddings. Entity type plays an important role only combined with other embeddings. Entity embedding improves \texttt{K-NRM} while has little effect on \texttt{Conv-KNRM}. This result further confirms that the signal from entity names are captured by the n-gram CNNs in \texttt{Conv-KNRM}. 
Incorporating all of three embeddings usually gets the best ranking performance.

This experiments shows that knowledge graph semantics are crucial to \texttt{EDRM}'s effectiveness. \texttt{Conv-KNRM} learns good phrase matches that overlap with the entity embedding signals. However, the knowledge graph semantics (descriptions and types) is hard to be learned just from user clicks.

\begin{figure}[t]
	\begin{minipage}[t]{0.48\textwidth}
		\centering
		\subfigure[K-NRM VS. EDRM]{
			\label{fig:difficulty:a}
			\includegraphics[width=0.48\textwidth]{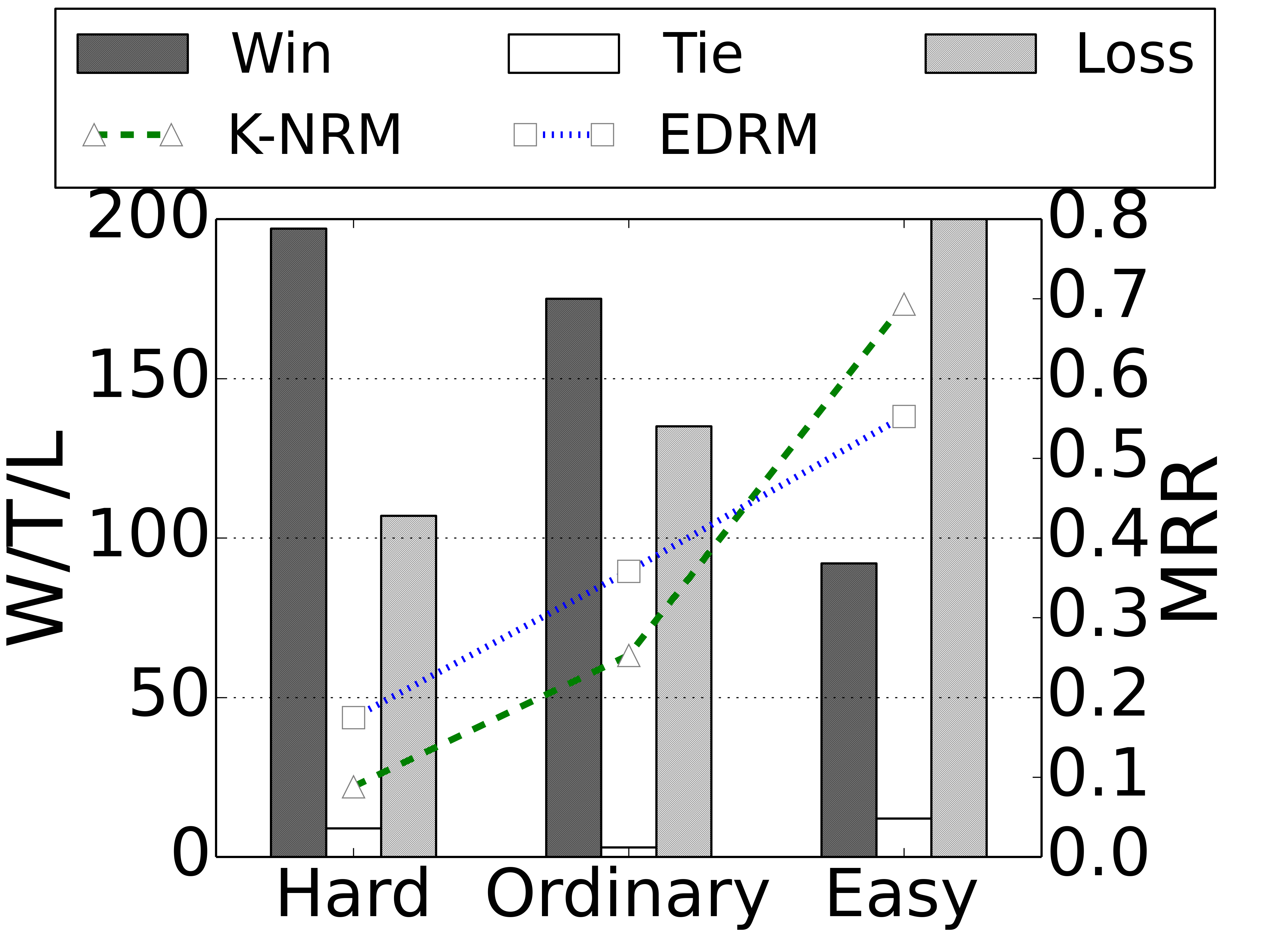}}
		\subfigure[Conv-KNRM VS. EDRM]{
			\label{fig:difficulty:b}
			\includegraphics[width=0.48\textwidth]{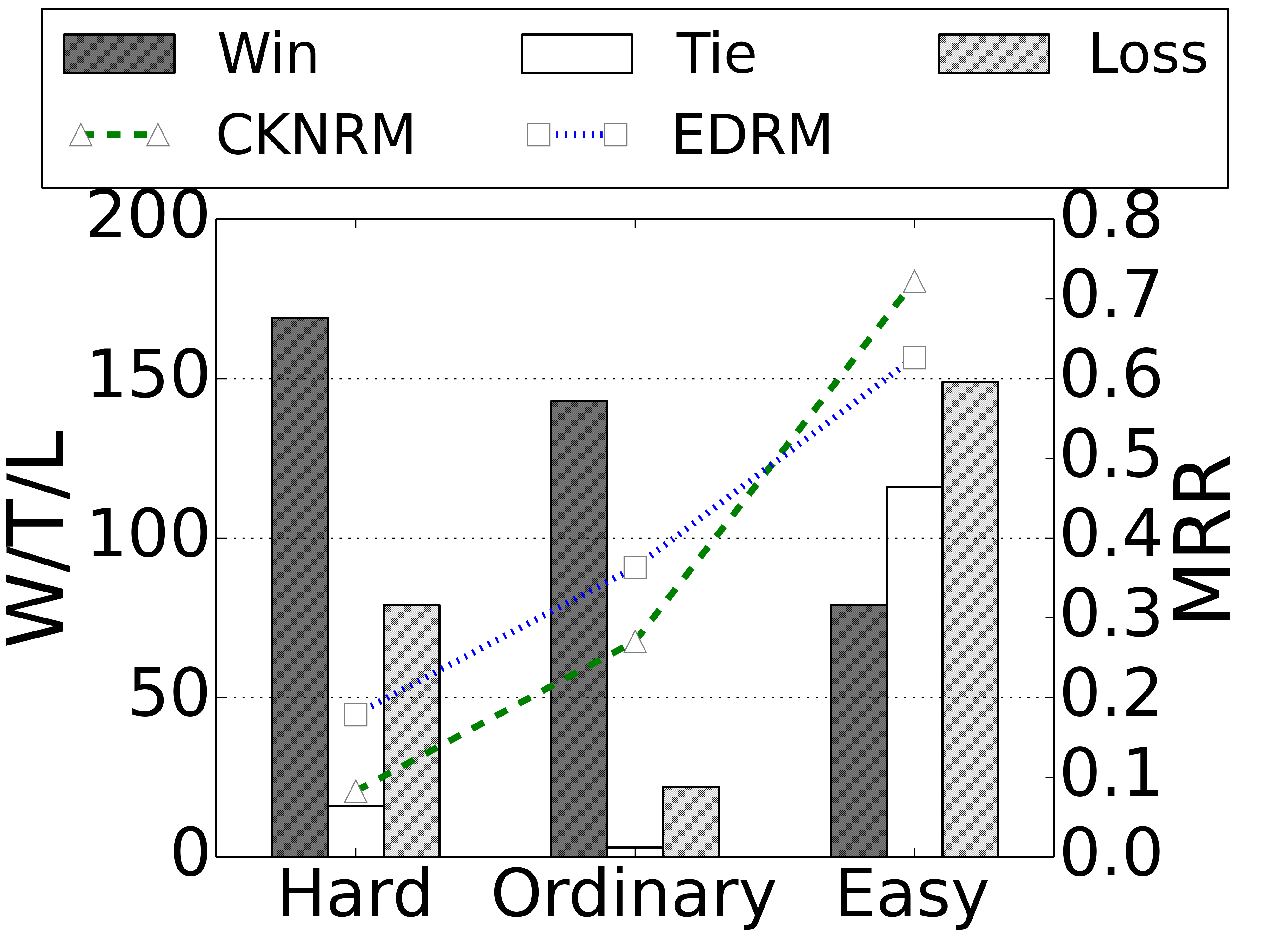}}
		\caption{Performance VS. Query Difficulty. The x-axises mark three query difficulty levels. The y-axises are the Win/Tie/Loss (left) and MRR (right) in the corresponding group.}\label{fig:difficulty}
	\end{minipage}

	\begin{minipage}[t]{0.48\textwidth}
		\centering
		\subfigure[K-NRM VS. EDRM]{
			\label{fig:length:a}
			\includegraphics[width=0.48\textwidth]{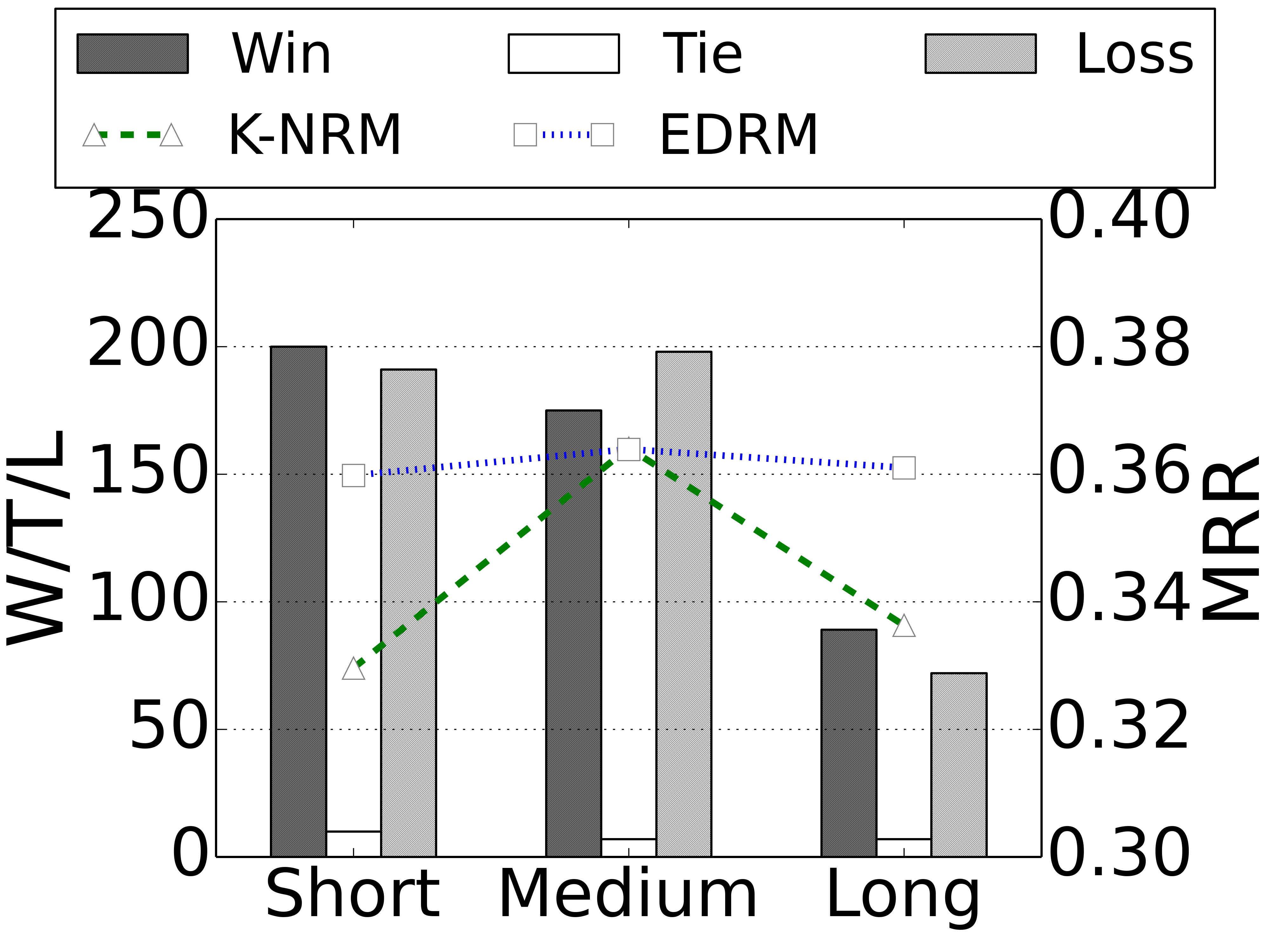}}
		\subfigure[Conv-KNRM VS. EDRM]{
			\label{fig:length:b}
			\includegraphics[width=0.48\textwidth]{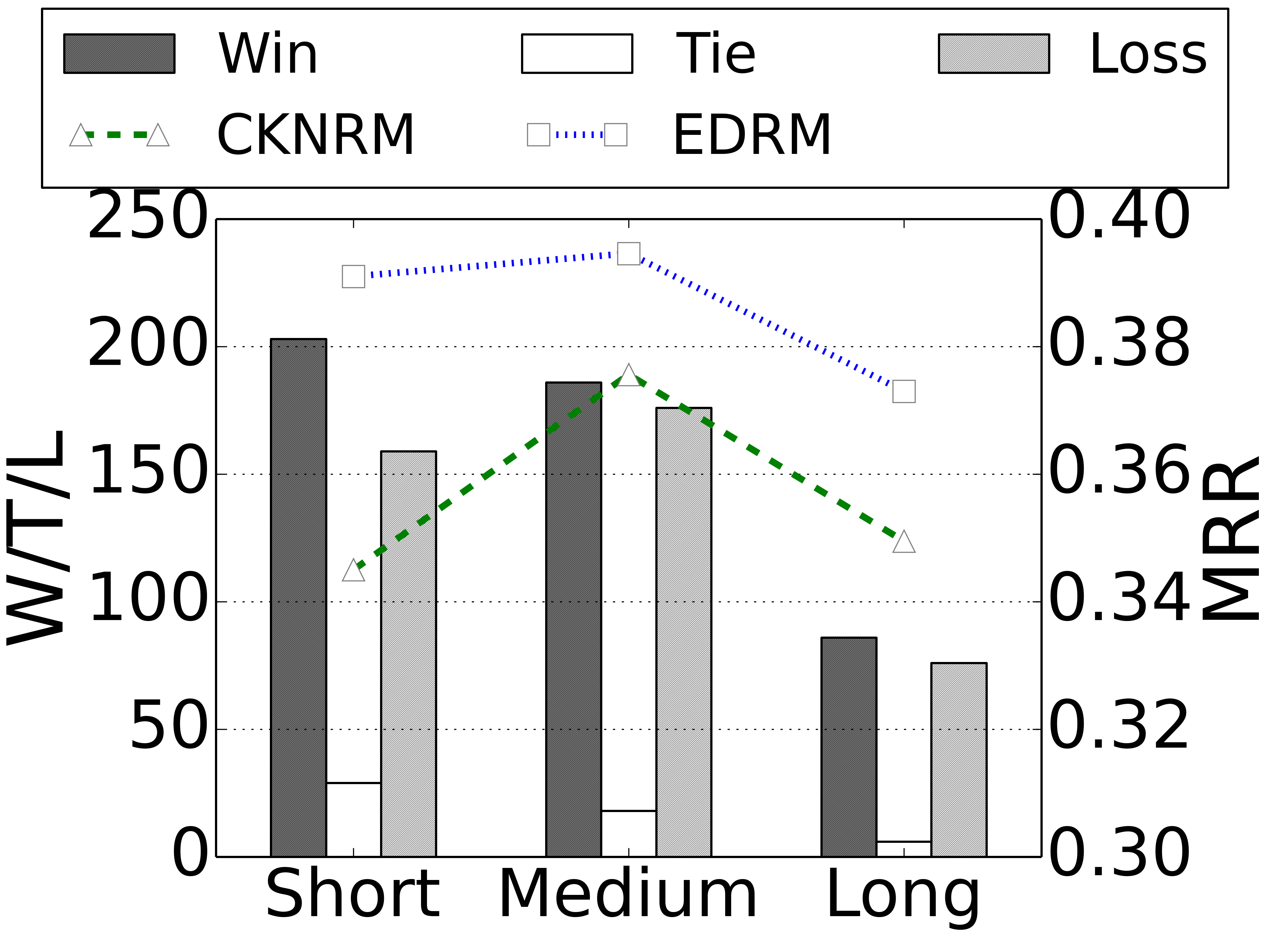}}
		\caption{Performance VS. Query Length. The x-axises mark three query length levels, and y-axises are the Win/Tie/Loss (left) and MRR (right) in the corresponding group.}\label{fig:length}
	\end{minipage}
\end{figure}

\subsection{Performance on Different Scenarios}
This experiment analyzes the influence of knowledge graphs in two different scenarios: multiple difficulty degrees and multiple length degrees.

\textbf{Query Difficulty Experiment} studies \texttt{EDRM}'s performance on testing queries at different difficulty, partitioned by \texttt{Conv-KNRM}'s MRR value: Hard (MRR $< 0.167$), Ordinary (MRR $\in [0.167, 0.382]$, and Easy (MRR $> 0.382$). As shown in Figure \ref{fig:difficulty}, \texttt{EDRM} performs the best on hard queries. 

\textbf{Query Length Experiment} evaluates \texttt{EDRM}'s effectiveness on Short (1 words), Medium (2-3 words) and Long (4 or more words) queries. As shown in Figure \ref{fig:length}, \texttt{EDRM} has more win cases and achieves the greatest improvement on short queries. Knowledge embeddings are more crucial when limited information is available from the original query text. 

These two experiments reveal that the effectiveness of \texttt{EDRM} is more observed on harder or shorter queries, whereas the word-based neural models either find it difficult or do not have sufficient information to leverage.

\begin{table*}[t]
	\centering
	\caption{Examples of entity semantics connecting query and title. All the examples are correctly ranked by EDRM-CKNRM. Table 3a shows query-document pairs. Table 3b lists the related entity semantics that include useful information to match the query-document pair. The examples and related semantics are picked by manually examining the ranking changes between different variances of \texttt{EDRM-CKNRM}.  \label{table.case}}
	\subtable[Query and document examples. \textit{Entities} are emphasized.]{
		\resizebox{\textwidth}{13mm}{
			\begin{tabular}{c|c}
				\hline
				\textbf{Query} & \textbf{Document} \\
				\hline 
				
				\textit{Meituxiuxiu web version} &\textit{Meituxiuxiu web version}: An online picture processing tools\\
				\hline 
				Home page of \textit{Meilishuo}& Home page of \textit{Meilishuo} - Only the correct popular fashion\\
				\hline 
				\textit{Master Lu} & Master Lu official website: \textit{System optimization}, hardware test, phone evaluation\\
				\hline 
				\textit{Crayon Shin-chan}: The movie & \textit{Crayon Shin-chan}: The movie online - Anime\\
				\hline 
				\textit{GINTAMA} & \textit{GINTAMA}: The movie online - Anime - Full HD online watch\\
				\hline 
				
	\end{tabular}}}
	
	\subtable[Semantics of related entities. The first two rows and last two rows show entity descriptions and entity types respectively.]{
		\resizebox{\textwidth}{15mm}{
			\begin{tabular}{c|c}
				\hline
				\textbf{Entity} & \textbf{Content}\\
				\hline
				\textit{Meituxiuxiu web version}&Description: Meituxiuxiu is the most popular Chinese image processing software,\\ 
				&launched by the Meitu company\\
				\hline
				\textit{Meilishuo}&Description: Meilishuo, the largest women's fashion e-commerce platform,\\
				&dedicates to provide the most popular fashion shopping experience\\
				\hline
				\textit{Crayon Shin-chan}, \textit{GINTAMA} & Type: Anime; Cartoon characters; Comic\\
				\hline
				
				\textit{Master Lu}, \textit{System Optimization} & Type: Hardware test; Software; System tool\\
				\hline
	\end{tabular}}}
\end{table*}
\subsection{Case Study}
Table~\ref{table.case} provide examples reflecting two possible ways, in which the knowledge graph semantics could help the document ranking.

First, the entity descriptions explain the meaning of entities and connect them through the word space. \textit{Meituxiuxiu web version} and \textit{Meilishuo} are two websites providing image processing and shopping services respectively. Their descriptions provide extra ranking signals to promote the related documents. 

Second, entity types establish underlying relevance patterns between query and documents. 
The underlying patterns can be captured by cross-space matches.
For example, the types of the query entity \textit{Crayon Shin-chan} and \textit{GINTAMA} overlaps with the bag-of-words in the relevant documents.
They can also be captured by the entity-based matches through their type overlaps, for example, between the query entity \textit{Master Lu} and the document entity \textit{System Optimization}.

\section{Conclusions}\label{section.conclusion}


This paper presents \texttt{EDRM}, the Entity-Duet Neural Ranking Model that incorporating knowledge graph semantics into neural ranking systems.
\texttt{EDRM} inherits entity-oriented search to match query and documents with bag-of-words and bag-of-entities in neural ranking models. 
The knowledge graph semantics are integrated as distributed representations of entities. The neural model leverages these semantics to help document ranking.
Using user clicks from search logs, the whole model---the integration of knowledge graph semantics and the neural ranking networks--is trained end-to-end.
It leads to a data-driven combination of entity-oriented search and neural information retrieval. 

Our experiments on the Sogou search log and CN-DBpedia demonstrate \texttt{EDRM}'s effectiveness and generalization ability over two state-of-the-art neural ranking models.
Our further analyses reveal that the generalization ability comes from the integration of knowledge graph semantics. 
The neural ranking models can effectively model n-gram matches between query and document, which overlaps with part of the ranking signals from entity-based matches: Solely adding the entity names may not improve the ranking accuracy much. 
However, the knowledge graph semantics, introduced by the description and type embeddings, provide novel ranking signals that greatly improve the generalization ability of neural rankers in difficult scenarios.

This paper preliminarily explores the role of structured semantics in deep learning models. Though mainly fouced on search, we hope our findings shed some lights on a potential path towards more intelligent neural systems and will motivate more explorations in this direction.

\section*{Acknowledgments}
This work\footnote{Source codes of this work are available at \\ \url{http://github.com/thunlp/EntityDuetNeuralRanking}} is supported by the Major Project of the National Social Science Foundation of China (No.13\&ZD190) as well as the China-Singapore Joint Research Project of the National Natural Science Foundation of China (No. 61661146007) under the umbrella of the NexT Joint Research Center of Tsinghua University and National University of Singapore. Chenyan Xiong is supported by National Science Foundation (NSF) grant IIS-1422676. We thank Sogou for
providing access to the search log.

\bibliographystyle{acl_natbib}

\newpage
\begin{thebibliography}{}
\expandafter\ifx\csname natexlab\endcsname\relax\def\natexlab#1{#1}\fi

\bibitem[{Auer et~al.(2007)Auer, Bizer, Kobilarov, Lehmann, Cyganiak, and
  Ives}]{auer2007dbpedia}
S{\"o}ren Auer, Christian Bizer, Georgi Kobilarov, Jens Lehmann, Richard
  Cyganiak, and Zachary Ives. 2007.
\newblock {\em {DBpedia}: A nucleus for a web of open data\/}.
\newblock Springer.

\bibitem[{Berger and Lafferty(1999)}]{berger1999Information}
Adam Berger and John Lafferty. 1999.
\newblock Information retrieval as statistical translation.
\newblock In {\em Proceedings of the 22nd Annual International ACM SIGIR
  Conference on Research and Development in Information Retrieval (SIGIR
  1999)\/}. ACM, pages 222--229.

\bibitem[{Bollacker et~al.(2008)Bollacker, Evans, Paritosh, Sturge, and
  Taylor}]{bollacker2008freebase}
Kurt Bollacker, Colin Evans, Praveen Paritosh, Tim Sturge, and Jamie Taylor.
  2008.
\newblock Freebase: A collaboratively created graph database for structuring
  human knowledge.
\newblock In {\em Proceedings of the 2008 ACM SIGMOD International Conference
  on Management of Data (SIGMOD 2008)\/}. ACM, pages 1247--1250.

\bibitem[{Cao et~al.(2008)Cao, Nie, Gao, and Robertson}]{cao2008selecting}
Guihong Cao, Jian-Yun Nie, Jianfeng Gao, and Stephen Robertson. 2008.
\newblock Selecting good expansion terms for pseudo-relevance feedback.
\newblock In {\em Proceedings of the 31st Annual International ACM SIGIR
  Conference on Research and Development in Information Retrieval (SIGIR
  2008)\/}. ACM, pages 243--250.

\bibitem[{Chuklin et~al.(2015)Chuklin, Markov, and Rijke}]{chuklin2015click}
Aleksandr Chuklin, Ilya Markov, and Maarten~de Rijke. 2015.
\newblock Click models for web search.
\newblock {\em Synthesis Lectures on Information Concepts, Retrieval, and
  Services\/} 7(3):1--115.

\bibitem[{Dai et~al.(2018)Dai, Xiong, Callan, and Liu}]{convknrm}
Zhuyun Dai, Chenyan Xiong, Jamie Callan, and Zhiyuan Liu. 2018.
\newblock Convolutional neural networks for soft-matching n-grams in ad-hoc
  search.
\newblock In {\em Proceedings of the Eleventh ACM International Conference on
  Web Search and Data Mining (WSDM 2018)\/}. ACM, pages 126--134.

\bibitem[{Dalton et~al.(2014)Dalton, Dietz, and Allan}]{daltonentity}
Jeffrey Dalton, Laura Dietz, and James Allan. 2014.
\newblock Entity query feature expansion using knowledge base links.
\newblock In {\em Proceedings of the 37th Annual International ACM SIGIR
  Conference on Research and Development in Information Retrieval (SIGIR
  2014)\/}. ACM, pages 365--374.

\bibitem[{Dehghani et~al.(2017)Dehghani, Zamani, Severyn, Kamps, and
  Croft}]{weeksupervise}
Mostafa Dehghani, Hamed Zamani, Aliaksei Severyn, Jaap Kamps, and W.~Bruce
  Croft. 2017.
\newblock Neural ranking models with weak supervision.
\newblock In {\em Proceedings of the 40th International ACM SIGIR Conference on
  Research and Development in Information Retrieval (SIGIR 2017)\/}. ACM, pages
  65--74.

\bibitem[{Dietz and Verga(2014)}]{daltontrec}
Laura Dietz and Patrick Verga. 2014.
\newblock Umass at {TREC} 2014: Entity query feature expansion using knowledge
  base links.
\newblock In {\em Proceedings of The 23st {Text} {Retrieval} {Conference} (TREC
  2014)\/}. NIST.

\bibitem[{Ensan and Bagheri(2017)}]{SELM}
Faezeh Ensan and Ebrahim Bagheri. 2017.
\newblock Document retrieval model through semantic linking.
\newblock In {\em Proceedings of the Tenth ACM International Conference on Web
  Search and Data Mining (WSDM 2017)\/}. ACM, pages 181--190.

\bibitem[{Ghazvininejad et~al.(2018)Ghazvininejad, Brockett, Chang, Dolan, Gao,
  Yih, and Galley}]{kgqa}
Marjan Ghazvininejad, Chris Brockett, Ming-Wei Chang, Bill Dolan, Jianfeng Gao,
  Scott Wen-tau Yih, and Michel Galley. 2018.
\newblock A knowledge-grounded neural conversation model.
\newblock In {\em The Thirty-Second AAAI Conference on Artificial Intelligence
  (AAAI 2018)\/}.

\bibitem[{Grauman and Darrell(2005)}]{grauman2005pyramid}
Kristen Grauman and Trevor Darrell. 2005.
\newblock The pyramid match kernel: Discriminative classification with sets of
  image features.
\newblock In {\em Tenth IEEE International Conference on Computer Vision
  (ICCV'05) Volume 1\/}. IEEE, volume~2, pages 1458--1465.

\bibitem[{Guo et~al.(2016{\natexlab{a}})Guo, Fan, Ai, and
  Croft}]{guo2016matching}
Jiafeng Guo, Yixing Fan, Qingyao Ai, and W.~Bruce Croft. 2016{\natexlab{a}}.
\newblock Semantic matching by non-linear word transportation for information
  retrieval.
\newblock In {\em Proceedings of the 25th ACM International on Conference on
  Information and Knowledge Management (CIKM 2016)\/}. ACM, pages 701--710.

\bibitem[{Guo et~al.(2016{\natexlab{b}})Guo, Fan, Ai, and
  Croft}]{jiafeng2016deep}
Jiafeng Guo, Yixing Fan, Qingyao Ai, and W.Bruce Croft. 2016{\natexlab{b}}.
\newblock A deep relevance matching model for ad-hoc retrieval.
\newblock In {\em Proceedings of the 25th ACM International on Conference on
  Information and Knowledge Management (CIKM 2016)\/}. ACM, pages 55--64.

\bibitem[{Gupta et~al.(2017)Gupta, Singh, and Roth}]{gupta2017entity}
Nitish Gupta, Sameer Singh, and Dan Roth. 2017.
\newblock Entity linking via joint encoding of types, descriptions, and
  context.
\newblock In {\em Proceedings of the 2017 Conference on Empirical Methods in
  Natural Language Processing (EMNLP 2017)\/}. pages 2681--2690.

\bibitem[{Hasibi et~al.(2017)Hasibi, Balog, and Bratsberg}]{hasibi2017entity}
Faegheh Hasibi, Krisztian Balog, and Svein~Erik Bratsberg. 2017.
\newblock Entity linking in queries: Efficiency vs. effectiveness.
\newblock In {\em European Conference on Information Retrieval\/}. Springer,
  pages 40--53.

\bibitem[{Hu et~al.(2014)Hu, Lu, Li, and Chen}]{arcii}
Baotian Hu, Zhengdong Lu, Hang Li, and Qingcai Chen. 2014.
\newblock Convolutional neural network architectures for matching natural
  language sentences.
\newblock In {\em Proceedings of the 27th International Conference on Neural
  Information Processing Systems - Volume 2 (NIPS 2014)\/}. MIT Press, pages
  2042--2050.

\bibitem[{Huang et~al.(2013)Huang, He, Gao, Deng, Acero, and
  Heck}]{huang2013learning}
Po-Sen Huang, Xiaodong He, Jianfeng Gao, Li~Deng, Alex Acero, and Larry Heck.
  2013.
\newblock Learning deep structured semantic models for web search using
  clickthrough data.
\newblock In {\em Proceedings of the 22nd ACM international conference on
  Conference on information \& knowledge management (CIKM 2013)\/}. ACM, pages
  2333--2338.

\bibitem[{Hui et~al.(2017)Hui, Yates, Berberich, and de~Melo}]{hui2017pacrr}
Kai Hui, Andrew Yates, Klaus Berberich, and Gerard de~Melo. 2017.
\newblock Pacrr: A position-aware neural ir model for relevance matching.
\newblock In {\em Proceedings of the 2017 Conference on Empirical Methods in
  Natural Language Processing (EMNLP 2017)\/}. pages 1060--1069.

\bibitem[{Joachims(2002)}]{ranksvm}
Thorsten Joachims. 2002.
\newblock Optimizing search engines using clickthrough data.
\newblock In {\em Proceedings of the 8th ACM SIGKDD International Conference on
  Knowledge Discovery and Data Mining (KDD 2002)\/}. ACM, pages 133--142.

\bibitem[{Liu and Fang(2015)}]{liu2015latent}
Xitong Liu and Hui Fang. 2015.
\newblock Latent entity space: {A} novel retrieval approach for entity-bearing
  queries.
\newblock {\em Information Retrieval Journal\/} 18(6):473--503.

\bibitem[{Luo et~al.(2017)Luo, Zheng, Liu, Wang, Xu, Zhang, and Ma}]{sogou}
Cheng Luo, Yukun Zheng, Yiqun Liu, Xiaochuan Wang, Jingfang Xu, Min Zhang, and
  Shaoping Ma. 2017.
\newblock Sogout-16: A new web corpus to embrace ir research.
\newblock In {\em Proceedings of the 40th International ACM SIGIR Conference on
  Research and Development in Information Retrieval (SIGIR 2017)\/}. ACM, pages
  1233--1236.

\bibitem[{Metzler and Croft(2006)}]{coorascent}
Donald Metzler and W.~Bruce Croft. 2006.
\newblock Linear feature-based models for information retrieval.
\newblock {\em Information Retrieval\/} 10(3):257--274.

\bibitem[{Miller et~al.(2016)Miller, Fisch, Dodge, Karimi, Bordes, and
  Weston}]{Miller2016KeyValueMN}
Alexander~H. Miller, Adam Fisch, Jesse Dodge, Amir-Hossein Karimi, Antoine
  Bordes, and Jason Weston. 2016.
\newblock Key-value memory networks for directly reading documents.
\newblock In {\em Proceedings of the 2016 Conference on Empirical Methods in
  Natural Language Processing (EMNLP 2016)\/}. pages 1400--1409.

\bibitem[{Mitra et~al.(2017)Mitra, Diaz, and Craswell}]{www2017duet}
Bhaskar Mitra, Fernando Diaz, and Nick Craswell. 2017.
\newblock Learning to match using local and distributed representations of text
  for web search.
\newblock In {\em Proceedings of the 26th International Conference on World
  Wide Web (WWW 2017)\/}. ACM, pages 1291--1299.

\bibitem[{Pang et~al.(2016)Pang, Lan, Guo, Xu, Wan, and Cheng}]{Pang2016TextMA}
Liang Pang, Yanyan Lan, Jiafeng Guo, Jun Xu, Shengxian Wan, and Xueqi Cheng.
  2016.
\newblock Text matching as image recognition.
\newblock In {\em In Proceedings of the Thirtieth AAAI Conference on Artificial
  Intelligence (AAAI 2016)\/}. pages 2793--2799.

\bibitem[{Pang et~al.(2017)Pang, Lan, Guo, Xu, Xu, and
  Cheng}]{pang2017deeprank}
Liang Pang, Yanyan Lan, Jiafeng Guo, Jun Xu, Jingfang Xu, and Xueqi Cheng.
  2017.
\newblock Deeprank: A new deep architecture for relevance ranking in
  information retrieval.
\newblock In {\em Proceedings of the 2017 ACM on Conference on Information and
  Knowledge Management (CIKM 2017)\/}. ACM, pages 257--266.

\bibitem[{Raviv et~al.(2016)Raviv, Kurland, and Carmel}]{raviv2016document}
Hadas Raviv, Oren Kurland, and David Carmel. 2016.
\newblock Document retrieval using entity-based language models.
\newblock In {\em Proceedings of the 39th Annual International ACM SIGIR
  Conference on Research and Development in Information Retrieval (SIGIR
  2016)\/}. ACM, pages 65--74.

\bibitem[{Shen et~al.(2014)Shen, He, Gao, Deng, and Mesnil}]{cdssm}
Yelong Shen, Xiaodong He, Jianfeng Gao, Li~Deng, and Gr{\'e}goire Mesnil. 2014.
\newblock A latent semantic model with convolutional-pooling structure for
  information retrieval.
\newblock In {\em Proceedings of the 23rd ACM International Conference on
  Conference on Information and Knowledge Management (CIKM 2014)\/}. ACM, pages
  101--110.

\bibitem[{Suchanek et~al.(2007)Suchanek, Kasneci, and
  Weikum}]{suchanek2007yago}
Fabian~M Suchanek, Gjergji Kasneci, and Gerhard Weikum. 2007.
\newblock Yago: a core of semantic knowledge.
\newblock In {\em Proceedings of the 16th international conference on World
  Wide Web (WWW 2007)\/}. ACM, pages 697--706.

\bibitem[{Wang et~al.(2013)Wang, Zhai, Dong, and Chang}]{tacm}
Hongning Wang, ChengXiang Zhai, Anlei Dong, and Yi~Chang. 2013.
\newblock Content-aware click modeling.
\newblock In {\em Proceedings of the 22Nd International Conference on World
  Wide Web (WWW 2013)\/}. ACM, pages 1365--1376.

\bibitem[{Xiong and Callan(2015)}]{EsdRank}
Chenyan Xiong and Jamie Callan. 2015.
\newblock {EsdRank}: Connecting query and documents through external
  semi-structured data.
\newblock In {\em Proceedings of the 24th ACM International on Conference on
  Information and Knowledge Management (CIKM 2015)\/}. ACM, pages 951--960.

\bibitem[{Xiong et~al.(2016)Xiong, Callan, and Liu}]{Xiong2016BOE}
Chenyan Xiong, Jamie Callan, and Tie-Yan Liu. 2016.
\newblock Bag-of-entities representation for ranking.
\newblock In {\em Proceedings of the sixth ACM International Conference on the
  Theory of Information Retrieval (ICTIR 2016)\/}. ACM, pages 181--184.

\bibitem[{Xiong et~al.(2017{\natexlab{a}})Xiong, Callan, and
  Liu}]{xiong2017duet}
Chenyan Xiong, Jamie Callan, and Tie-Yan Liu. 2017{\natexlab{a}}.
\newblock Word-entity duet representations for document ranking.
\newblock In {\em Proceedings of the 40th Annual International ACM SIGIR
  Conference on Research and Development in Information Retrieval (SIGIR
  2017)\/}. ACM, pages 763--772.

\bibitem[{Xiong et~al.(2017{\natexlab{b}})Xiong, Dai, Callan, Liu, and
  Power}]{xiong2017knrm}
Chenyan Xiong, Zhuyun Dai, Jamie Callan, Zhiyuan Liu, and Russell Power.
  2017{\natexlab{b}}.
\newblock End-to-end neural ad-hoc ranking with kernel pooling.
\newblock In {\em Proceedings of the 40th annual international ACM SIGIR
  conference on Research and Development in Information Retrieval (SIGIR
  2017)\/}. ACM, pages 55--64.

\bibitem[{Xiong et~al.(2017{\natexlab{c}})Xiong, Power, and Callan}]{ESR}
Chenyan Xiong, Russell Power, and Jamie Callan. 2017{\natexlab{c}}.
\newblock Explicit semantic ranking for academic search via knowledge graph
  embedding.
\newblock In {\em Proceedings of the 26th International Conference on World
  Wide Web (WWW 2017)\/}. ACM, pages 1271--1279.

\bibitem[{Xu et~al.(2017)Xu, Xu, Liang, Xie, Liang, Cui, and Xiao}]{xu2017cn}
Bo~Xu, Yong Xu, Jiaqing Liang, Chenhao Xie, Bin Liang, Wanyun Cui, and Yanghua
  Xiao. 2017.
\newblock Cn-dbpedia: A never-ending chinese knowledge extraction system.
\newblock In {\em International Conference on Industrial, Engineering and Other
  Applications of Applied Intelligent Systems\/}. Springer, pages 428--438.

\bibitem[{Zhang et~al.(2003)Zhang, Yu, Xiong, and Liu}]{ICTCLAS}
Hua~Ping Zhang, Hong~Kui Yu, De~Yi Xiong, and Qun Liu. 2003.
\newblock Hhmm-based chinese lexical analyzer ictclas.
\newblock In {\em Sighan Workshop on Chinese Language Processing\/}. pages
  758--759.

\end{thebibliography}

\appendix
\end{document}